\documentclass[showpacs,prb,twocolumn,superscriptaddress]{revtex4}

\usepackage[english]{babel}
\usepackage{amssymb,amsmath}
\usepackage{dsfont}
\usepackage{gensymb}
\usepackage{color}
\usepackage{epsfig}
\usepackage{graphicx}
\usepackage{ulem}

\begin{document}

\title{Combined effect of strain and defects on the conductance of graphene nanoribbons}

\author{Thomas~Lehmann}
\affiliation{Institute for Materials Science and Max Bergmann Center of Biomaterials, TU Dresden, 01062 Dresden, Germany}
\author{Dmitry\,A.~Ryndyk}
\affiliation{Institute for Materials Science and Max Bergmann Center of Biomaterials, TU Dresden, 01062 Dresden, Germany}
\affiliation{Center for Advancing Electronics Dresden, TU Dresden, 01062 Dresden, Germany}
\author{Gianaurelio~Cuniberti}
\affiliation{Institute for Materials Science and Max Bergmann Center of Biomaterials, TU Dresden, 01062 Dresden, Germany}
\affiliation{Center for Advancing Electronics Dresden, TU Dresden, 01062 Dresden, Germany}
\affiliation{Division of IT Convergence Engineering, POSTECH, Pohang 790-784, Republic of Korea}

\date{\today}

\begin{abstract}
We investigate the combined influence of structural defects and uniaxial longitudinal strain on the electronic transport properties of armchair graphene nanoribbons using the numerical approach based on the semi-empirical tight-binding model, the Landauer formalism and the recursion method for Green functions. We calculate the conductance of graphene nanoribbons in the quantum coherent regime with different types and concentrations of defects. Further, we apply uniform planar tension to the non-ideal graphene ribbons with randomly distributed and oriented single and double vacancies and Stone-Wales defects. Since transport characteristics of graphene nanoribbons are found to be very sensitive to edge termination and aspect ratio and it has been shown that energy gaps can emerge under critical strain, the interplay of both effects needs to be studied. We show that band gap engineering using strain is still possible for non-ideal armchair ribbons with a small defect concentration, as the oscillatory behaviour of the gap is preserved.
\end{abstract}

\pacs{\vspace{-0.1cm} 73.63.-b, 72.80.Vp}

\maketitle

\section{Introduction}

\begin{figure}[t]
\begin{center}
\raggedright
(a)\\
\vspace{-0.4cm}
\includegraphics[width=1\hsize]{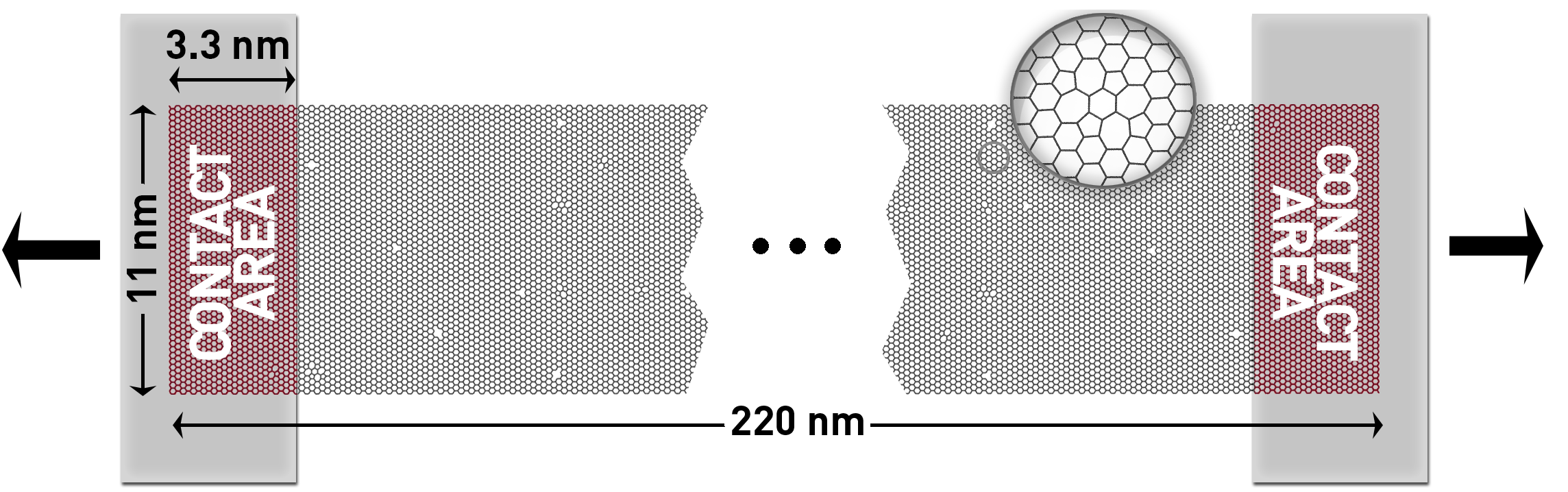}
\raggedright
(b)\\
\hspace{0.53cm}
\epsfxsize=0.926\hsize
\epsfbox{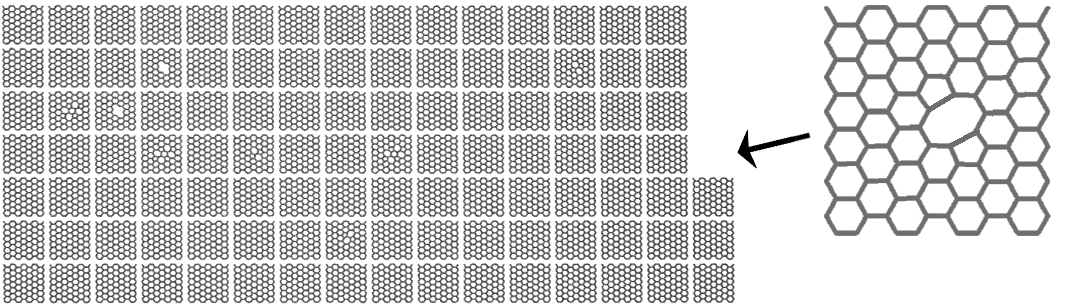}
\caption{(a) Conceptual sketch of the model setup. Longitudinal tensile uniaxial strain is applied to a defected armchair graphene nanoribbon, which is contacted to metallic wide-band electrodes at each side. (b) Visualization of the modular approach to construct a large defected ribbon from small geometry optimized segments.}
\label{fig:sketch}
\end{center}
\end{figure}

Since its first isolation by Novosolev and Geim\cite{Novoselov:2004} in 2004, graphene has been under intensive studies. Recently, tremendous interest in the use for electronic nanodevices has been raised because of its outstanding features, especially the ballistic transport and high electron mobility at room temperatures.

Graphene with its zero-gap property is not applicable for logic electronics. Therefore, semi-conducting graphene nanoribbons (GNR) are in high demand for many applications in future nanodevices, raising the question how to effectively tune the energy gap. Besides constraining the ribbon width to several nanometers or doping, mechanical deformation is a suitable candidate. The effect of uniaxial mechanical strain in ideal GNRs has been adressed by several theoretical studies \cite{Lu:2010fj,Faccio:2009,Sun:2008qy,Li:2010,Pereira:2009,Rosenkranz:2011,Ni:2008rt,Lu:2012,Poetschke:2010,Erdogan:2011} before and there is consensus that the spectral band gap is indeed sensitive to external strain. For zigzag graphene nanoribbons (ZGNR) one observes a monotonous increase for tensile uniaxial strain, whereas for armchair terminated ribbons (AGNR) the situation becomes more involved. AGNRs display a triangular oscillating feature, with a width dependent periodicity and amplitude. The two substantially different observations can be related to the totally distinct underlying mechanisms. For armchair termination the oscillatory band gap emerges due to shifting of the Fermi point perpendicular to the $\mathbf{k}$ values of allowed electronic states forming parallel lines.\cite{Yang:2000} Whereas in ZGNRs the band gap widening is caused by strain improved spin polarization in the edge states.\cite{Lu:2010fj} 

In this paper, we deepen the understanding of electronic properties of non-ideal strained GNR by studying the combined effect of uniaxial strain and structural defects on the conductance properties. We combine \textit{ab initio} structure studies with a semi-empirical tight-binding model for large scale electron transport calculations in the quantum coherent regime of a ribbon, strongly coupled to metallic electrodes in the wide-band limit. 

As all previous studies assumed the ribbons to be ideal, we analyze the effect of uniaxial strain in a more realistic setup with reference to potential applications. We include the existence of several vacancy defect types with various concentrations and adopt an improved contact model in contrast to the widely used semi-infinite graphene leads.
In addition to the case of straining the system on purpose for band gap engineering, there are also many circumstances where strain is not intentional but can not be avoided, e.g. due to lattice mismatch with the substrate or contacting to electrodes. Therefore, the understanding of the interplay between strain and defects in non-ideal ribbons gets even more important. 
We will only focus on AGNRs, because its band gap is more sensitive to strain resulting in better tuning capabilities and the fact that in ZGNRs the transport properties are mainly governed by edge states, requiring spin polarized transport calculations.

The paper is organized as follows. In Sec.\,\ref{sec:model} we briefly introduce our model, in particular we explain our modular approach for the ribbon geometry and provide details on the contact type, defect varieties and mechanical straining. Furthermore, we explain our used methods for geometry optimization and transport calculations. Subsequently we present our results on mechanical properties and transport characteristics in Sec.\,\ref{sec:results} and discuss the combined effect of strain and defects on the conductance properties of AGNRs. We conclude in Sec.\,\ref{sec:conclusions}.


\section{Model and Method}
\label{sec:model}

\begin{figure}
\begin{center}
\epsfxsize=1\hsize
\epsfbox{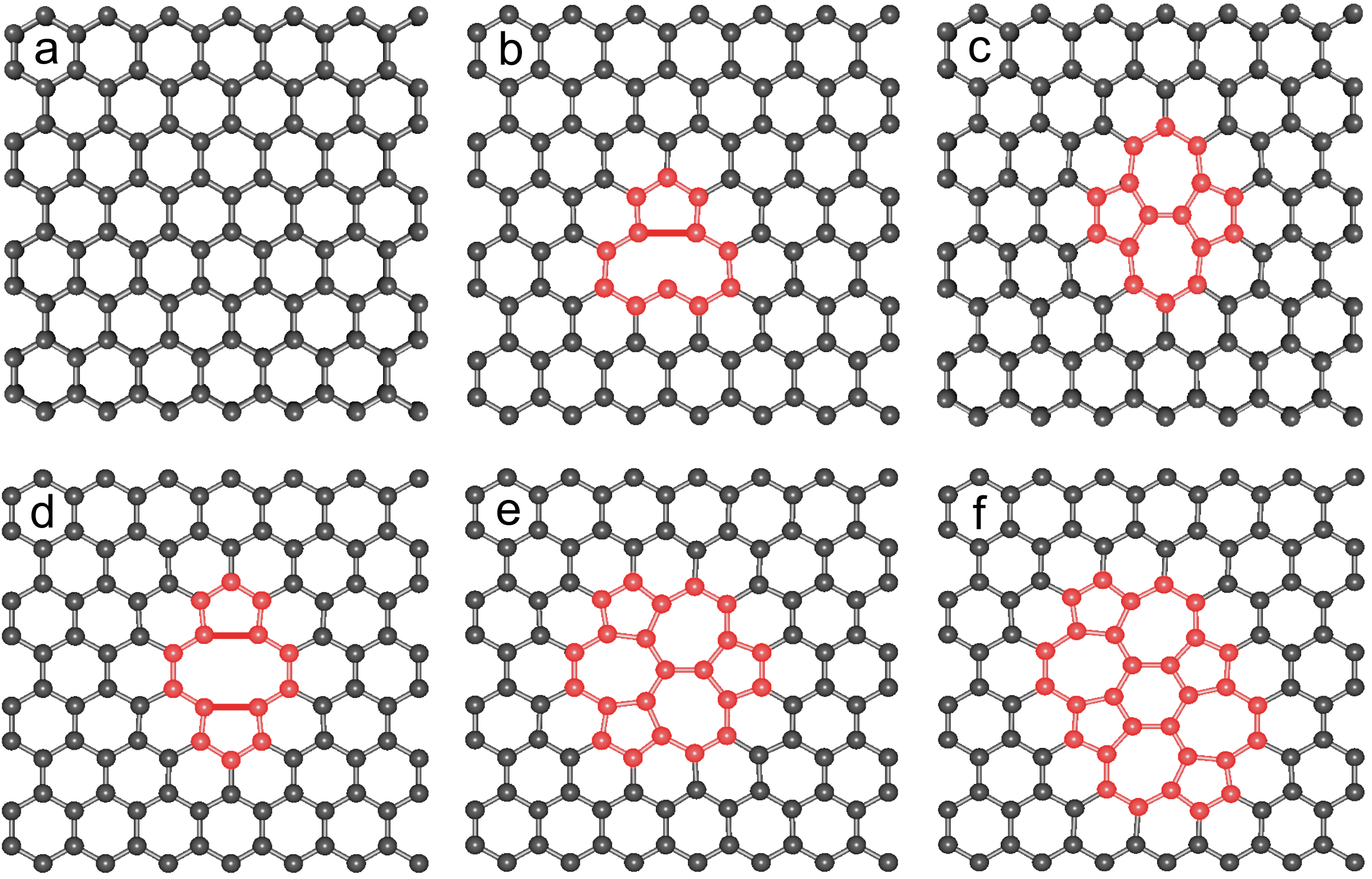}
\caption{(Color online) The six various types of building blocks, which were used to construct a large defected GNR: (a) pure, (b) single vacancy $V_1($5-9), (c)  Stone-Wales $SW$(55-77), (d) double vacancy $V_2($5-8-5),  (e) double vacancy $V_2$(555-777), (f) double vacancy $V_2($5555-6-7777). Additionally, all possible defect orientations have been considered.}
\label{fig:defects}
\end{center}
\end{figure}

Throughout this work we have used a long and narrow armchair graphene nanoribbon, which consists of 1120 zigzag rows in longitudinal direction with 98 atoms each, resulting in dimensions of approximately 11 nm in width and 220 nm in length.
This high aspect ratio ribbon is side-contacted to metallic wide-band leads, with a contacted length of about 3.3 nm, i.e. 16 zigzag rows, at each end. We include several defect types, such as Stone-Wales, single and double vacancies, randomly distributed and oriented to account for a non-ideal, more realistic geometry (see Fig.\,\ref{fig:sketch}a), but the contact areas remain undefected. We want to point out, that Stone-Wales defects are actually no vacancy defects, but just 90\degree-rotated carbon-carbon bonds producing two pentagon-heptagon pairs. However, as it has been often identified in TEM images, we included it here. For the double vacancies, there exist actually three different configurations: $V_2$(5-8-5), $V_2$(555-777) and $V_2$(5555-6-7777), all emerging from each other by bond rotation. As experimental values for the concentrations of the individual defect types are still missing, we use equal probabilities for all varieties. 
To set up our system geometry, we start from a modular approach. Using various kinds of small and equally sized graphene patches, which include one of the defect types described above (see Fig.\,\ref{fig:defects}) and are uniaxially strained by a certain value, we construct a large, high aspect-ratio ribbon. This technique allows us to investigate much larger defected structures, not being constrained by computational intense geometry optimizations. 

For the mechanical straining, we obtain the strained lattice vectors by performing structural optimization of atomic positions and supercell vectors of an infinite graphene sheet with external stress using first-principle density functional theory, as implemented in the \texttt{SIESTA} package.\cite{Soler:2002} We employ Troullier-Martins pseudopotentials to treat valence electrons with Perdew-Burke-Ernzerhof (PBE) exchange-correlation functionals in the general gradient approximation (GGA) using a double-zeta basis set with polarization orbitals. A real-space grid equivalent to an energy cutoff of 400 Ry and a Monkhorst-Pack grid of 15x15x1 has been set up. For the self-consistent calculation, we chose an energy tolerance of $10^{-5}$ eV, force tolerance of $0.01$ eV/A, stress tolerance of $0.1$ GPa and a target stress tensor in tensile, longitudinal direction of various strengths, up to 14 GPa. The largest strain value observed was about 15\% and we concentrated only on elongating deformations, longitudinal compression has not been considered here but the qualitative picture should be the same.

Using those relaxed lattice vectors, we construct small graphene patches, consisting of 8 zigzag rows with 14 atoms each, where we inserted one of the above mentioned defect types. For each of those individual fragments we performed again a geometry optimization with similar settings, but without external stress tensor and no $\mathbf{k}$-sampling, as the system is reasonably large. We fixed the atomic positions of boundary atoms, namely the outmost zigzag line and the first two dimer lines, to allow for a seamless joining of the segments. The patch size has been chosen large enough that the boundary conditions do not perturb the defect. The relaxed defect geometries strongly resemble structures observed in TEM images,\cite{Meyer:2008} confirming the structural results. Finally, nearly one thousand of randomly selected fragments, pure and defected, were used to construct the system of about 220 nm in length. The influence of edge vacancies has not been studied in particular, but edge effects do not play any significant role in AGNRs compared to ZGNRs. Further it has been shown, that in AGNRs transmission is nearly independent by the depth of edge defects,\cite{Dietl} verifying the validity of our approach.

To investigate the electronic transport properties of this system in the quantum coherent regime, we mapped the structure to an one-orbital tight-binding Hamiltonian to describe noninteracting $\pi$-electrons, considering only nearest neighbor hopping:
\begin{equation}
  H = \varepsilon_0 \sum_i c_i^\dag c_i^{\phantom{\dag}}  + \sum_{\langle i,j\rangle}  t(|\vec{r}_{ij}|) \, c_i^\dag c_j^{\phantom{\dag}}. \label{eq:H}
\end{equation}
Here $\varepsilon_0$ is the on-site energy, which gives only an energy offset and thus can be set to zero, $ c_i $ and $ c_i^\dag$ are the creation and annihilation operators for an electron in the $p_z$ orbital centered on the $i$-th carbon atom and $\langle i,j\rangle$ indicates all pairs of nearest neighbors. We assume a distance dependent hopping parameter $t$ with an exponential decay\cite{Pereira:2009}
\begin{equation}
  t(|\vec{r}_{ij}|) = t_0 \; exp\left(-3.37\left(\frac{|\vec{r}_{ij}|}{a}-1\right)\right), \label{eq:t}
\end{equation}
where $|\vec{r}_{ij}|$ is the distance between two neighboring atoms, $a=1.428$ \AA\, the equilibrium carbon-carbon bond length and $t_0=2.7$ eV the hopping in equilibrated graphene.\cite{Castro-Neto:2009} In an atomic basis, $\hat H$ becomes a block-tridiagonal matrix and the usual lead-system-lead tri-partitioning scheme can be applied.

A nearest-neighbor approximation has been chosen for simplicity and is found to describe low energy properties good enough for many applications.
The transport properties can now be obtained by using the standard Green function formalism. \cite{Datta95book,Ferry:2009,Cuevas10book,Ryndyk09inbook} The retarded Green function of the system is determined by the expression
\begin{equation}
  \hat G^\textrm{R} (E) =\left[E\hat S - \hat H - \hat\Sigma_{\mathsf{L}} -\hat\Sigma_{\mathrm{R}} \right]^{-1}, \label{eq:Gs}
\end{equation}
with the hermitian Hamiltonian $H_{ij}=\langle i| H|j \rangle$, energy $E$, overlap matrix $\hat S$ which is defined as the overlap integral $S_{ij}=\langle i|j \rangle$ and the left and right self-energies $\hat \Sigma_\alpha$ ($\alpha=\mathrm{L},\mathrm{R}$) to include coupling to the leads. Note that in our case the basis is orthogonal, i.e. $S_{ij}=\delta_{ij}$. The transmission function of the system 
\begin{equation}
  T(E) = \textrm{Tr}(\hat\Gamma_{\mathrm{L}} \hat G^{\textrm{R}} \hat \Gamma_{\mathrm{R}} \hat G^{\textrm{A}}) \label{eq:Ts}
\end{equation}
can then be calculated by Eq.\,(\ref{eq:Ts}) with the left and right broadening functions $\hat\Gamma_\alpha=-2\,\textrm{Im}\hat\Sigma_\alpha$. Using the Landauer formalism, the conductance is then proportional to $T(E)$ with the conductance quantum $G_0$ as prefactor with spin taken into account:
\begin{equation}
  G(E) = \frac{2e^2}{h} T(E). \label{eq:Con}
\end{equation}
The density of states can also be expressed in terms of $\hat G^\textrm{R}$:
\begin{equation}
  DOS(E) = - \frac{1}{\pi}  \textrm{Im}( \textrm{Tr}\;\hat G^\textrm{R}). \label{eq:DOSs}
\end{equation}

The direct calculation of the retarded Green function by means of matrix inversion as indicated in Eq.\,(\ref{eq:Gs}) becomes computationally intractable for large systems. To bypass this bottleneck, we exploit the recursive Green function technique, which is based on the Dyson equation and allows a very fast calculation of $\hat G^\textrm{R}$ especially for high aspect-ratio systems, by cutting the system in small slices and building it up slice by slice. Note that this slicing is completely independent from our modular construction of the ribbon geometry. This recursive approach speeds up the calculation from third order $O(N^3)$  in number of atoms $N$ to a first order scaling $O(NM^3)$ with $M$ being the number of states per slice. 
One gets the system of recursive equations:\footnote{For the sake of clarity we introduce a new notation. As all Green functions in this algorithm are retarded Green functions, the retarded/advanced index has been dropped. Instead, the upper index denotes the system described by the Green function, i.e. how many slices have been connected beginning from the left lead ($\rightarrow$) or right lead ($\leftarrow$). The lower index pair indicates the two system slices between which the propagator is calculated.}

\begin{align} 
 \nonumber
 \hat G_{i+1,i+1}^{\rightarrow i+1} &=\left[ E\hat S - \hat H_{i+1,i+1}^{\phantom{\rightarrow i}} - \hat\Sigma_{\mathrm{L}}\delta_{i,0} - \hat\Sigma_{\mathrm{R}}\delta_{i,N-1} \right.  \\
    &\textrm{ ~~~~}\displaystyle \left.- (E\hat S-\hat H_{i+1,i}^{\phantom{\rightarrow i}}) \hat  G_{i,i}^{\rightarrow i}(E\hat S-\hat H_{i,i+1}^{\phantom{\rightarrow i}}) \right]^{-1} \label{eq:rec1}\\
\hat G_{1,i+1}^{\rightarrow i+1} &= \hat G_{1,i}^{\rightarrow i} (E\hat S-\hat H_{i,i+1}^{\phantom{\rightarrow i+1}})  \hat G_{i+1,i+1}^{\rightarrow i+1}.  \label{eq:rec2}
\end{align}

Here, $\hat G_{1,i}^{\rightarrow i}$ is the propagator between the first and the $i$-th slice of the system connected up to the $i$-th slice to the left lead and $\hat H_{i,i+1}^{\phantom{\rightarrow i+1}}$ is the coupling of slice $i$ and $(i+1)$. The effect of the leads is again described by the self-energies $\hat\Sigma_\alpha$, which have to be considered only for the first and last slice. 
$\hat G^{\rightarrow N}_{1,N}$ is sufficient to calculate the transmission function, as can be seen in Eq.\,(\ref{eq:T}). Additional computational effort, called backward recursion, is required for the density of states, see Eq.\,(\ref{eq:DOS}), as one needs all diagonal elements $\hat G^{\rightarrow N}_{i,i}$, but it scales with $O(NM^2)$ faster than Eq.\,(\ref{eq:rec1}) - (\ref{eq:rec2}).

\begin{align}
  T(E) &= \textrm{Tr}(\hat\Gamma_{\mathrm{L}} \hat G^{\rightarrow N}_{1,N}\hat\Gamma_{\mathrm{R}} \hat G^{\rightarrow N \dag}_{1,N}) \label{eq:T}\\
  DOS(E) &= - \frac{1}{\pi} \textrm{Im}( \sum_{i=1}^N \textrm{Tr}\,\hat G^{\rightarrow N}_{i,i}) \label{eq:DOS}
\end{align}

For a detailed description of this algorithm we refer to the respective literature.\cite{MacKinnon:1985,Thouless:1981a,LEE:1981}

\section{Results}
\label{sec:results}

\begin{figure}[b]
\begin{center}
\raggedright
\epsfxsize=.95\hsize
\epsfbox{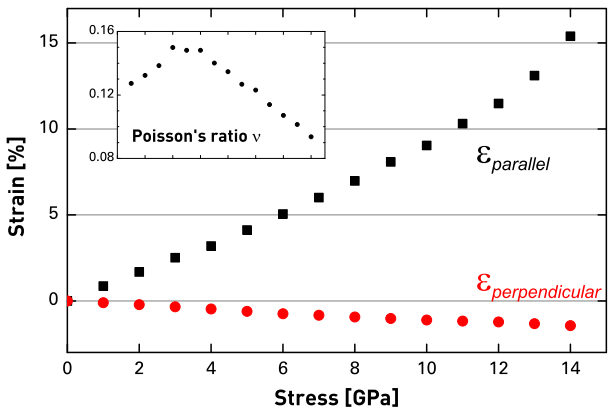}
\caption{(Color online) Stress-strain curve for an infinite graphene sheet as obtained by density-functional theory. As expected, a longitudinal tensile uniaxial strain $\varepsilon_\parallel$ leads to compression in transverse direction $\varepsilon_\perp$, resulting in a positive Poisson's ratio $\nu$ which values are consistent with other ab-initio studies (see inset).\cite{Kudin:2001,Reddy:2006,Faccio:2009}}
\label{fig:stress-strain}
\end{center}
\end{figure}

\begin{figure}
\raggedright
(a)
\epsfxsize=1\hsize
\epsfbox{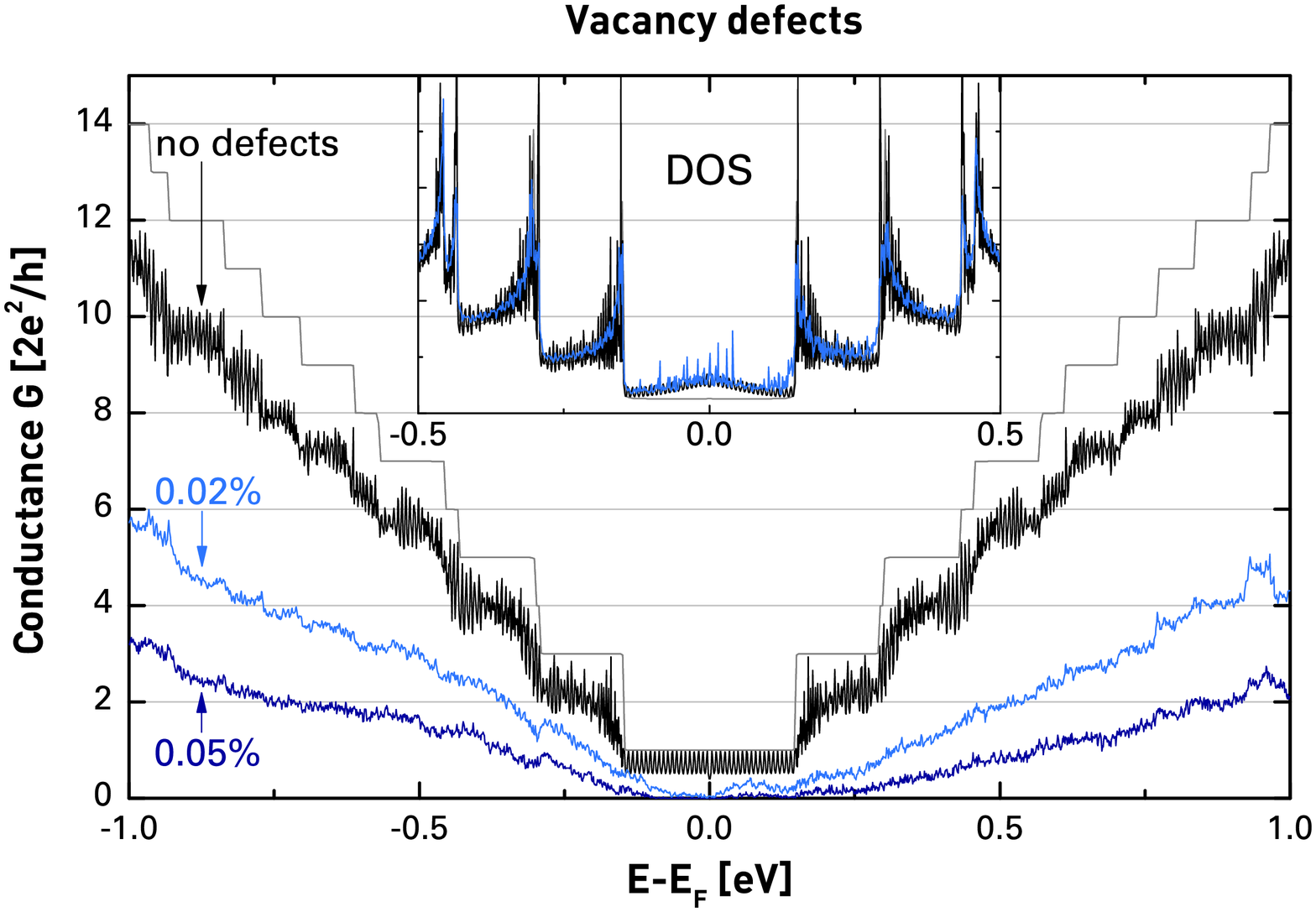}
(b)
\epsfxsize=1\hsize
\epsfbox{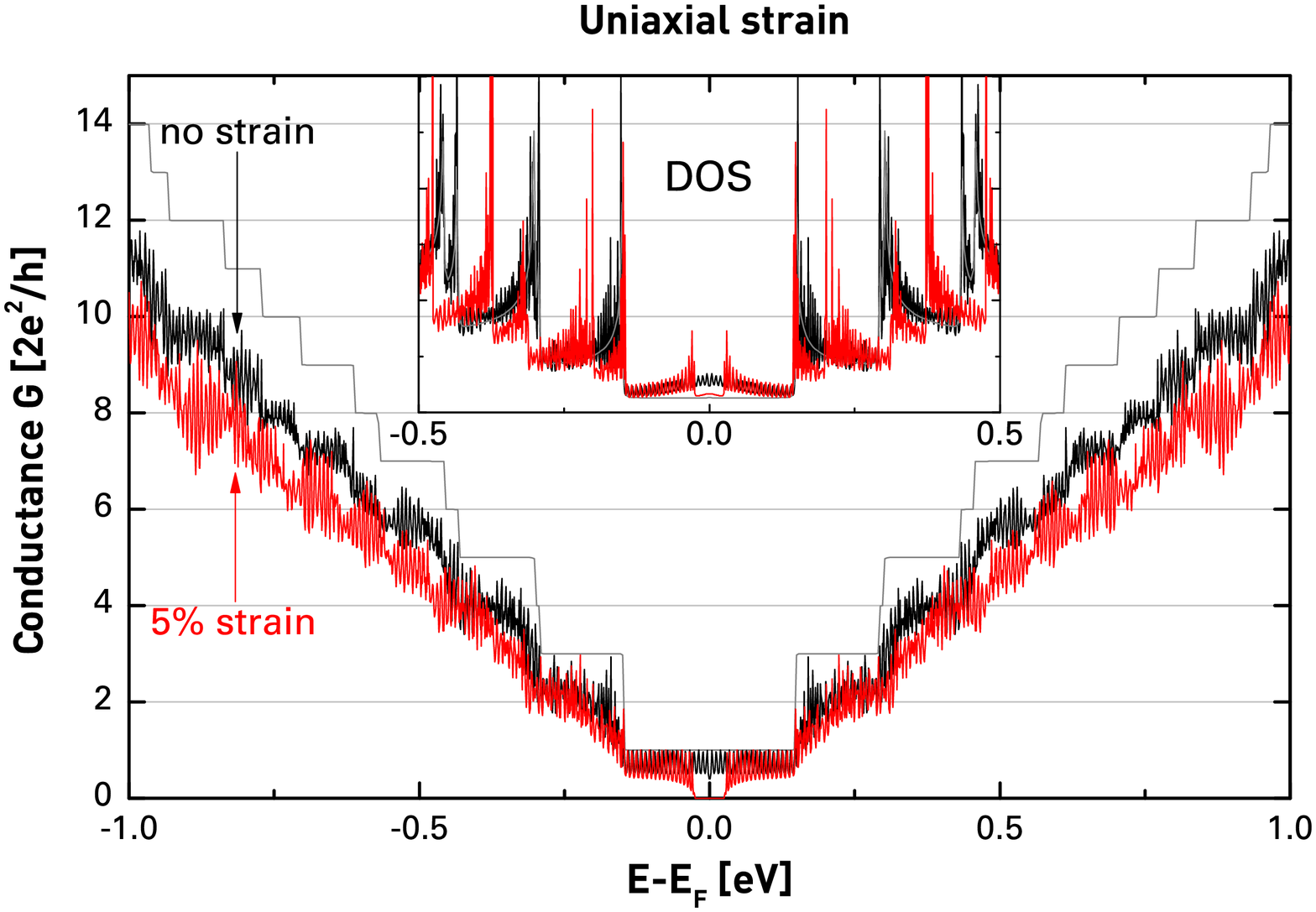}
(c)
\epsfxsize=1\hsize
\epsfbox{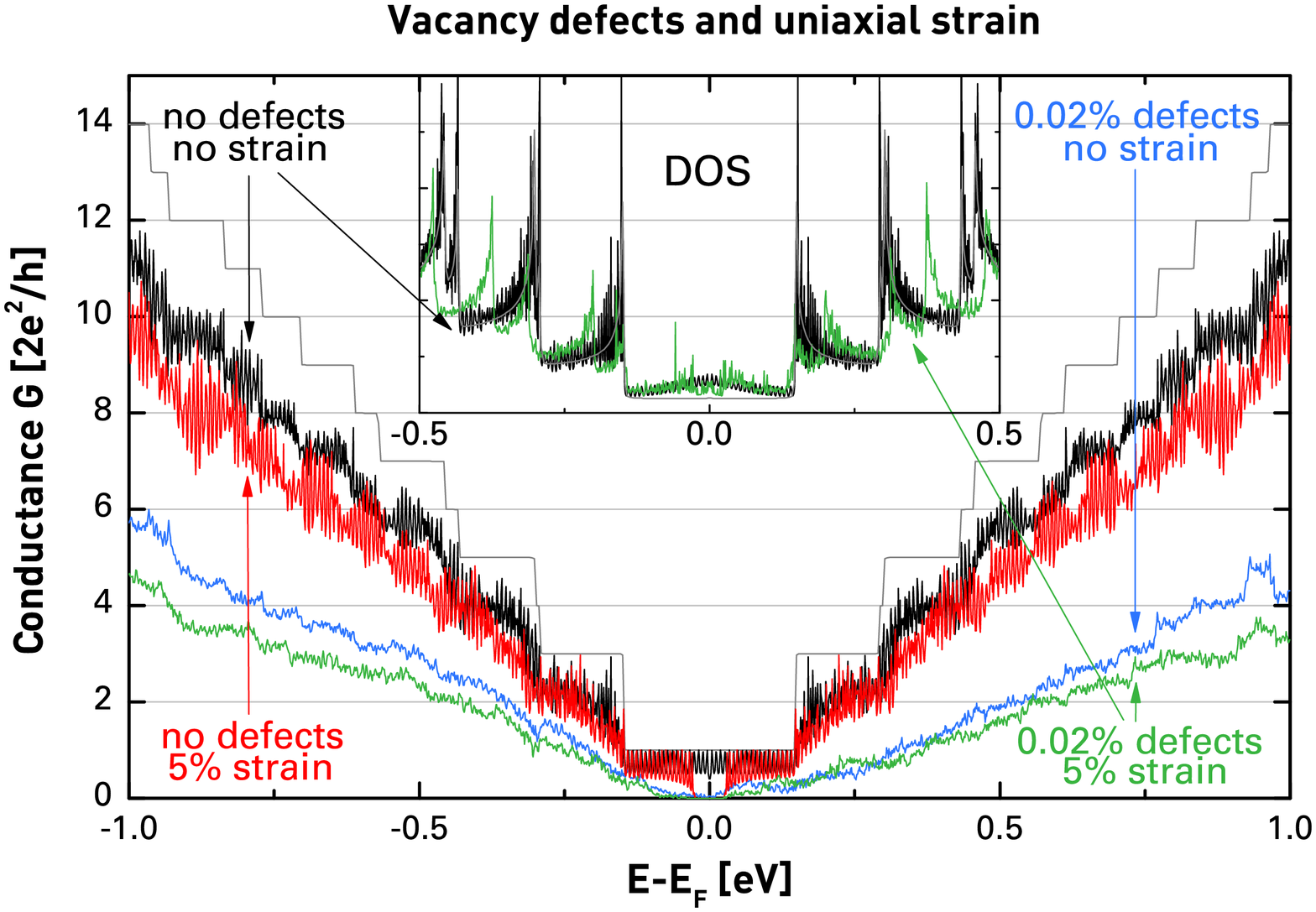}
\caption{(Color online) Conductance spectra for the 98-AGNR with (a) randomly distributed vacancy defects and zero strain, (b) longitudinal uniaxial strain without defects and (c) both, vacancy defects and uniaxial strain combined. The step-wise behaviour of the ribbon with semi-infinite leads is shown on top for reference, which gives an upper quantum limit for the conductance. The spectrum for the non-defected unstrained ribbon with side-contacted metal electrodes is also shown in black for comparison, while the effect of strain and defects can be seen in the colored spectra. All data sets for defected ribbons have been averaged over 20 equivalent samples. The density of states (DOS) is presented as an inset to show the effect on the DOS of the ideal ribbon for each of the three cases. }
\label{fig:str}
\end{figure}

\begin{figure}
\begin{center}
\epsfxsize=1\hsize
\epsfbox{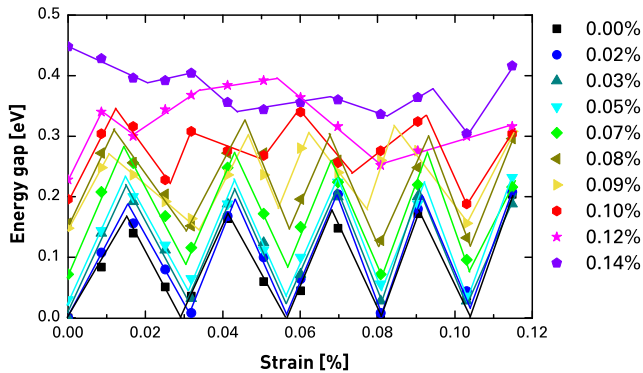}
\caption{(Color online) Strain dependent energy gap in the conductance for different defect concentrations, illustrating the combined effect of uniaxial strain and vacancy defects on the conductance properties of AGNRs. Oscillation is well preserved for small defect concentrations, but being suppressed for high enough values and saturates at a certain level for concentrations of about $0.1\%$ or higher.}
\label{fig:gap}
\end{center}
\end{figure}

Armchair type ribbons can be grouped into three families according to their width. They are classified by $N_a=3m$, $3m+1$ or $3m+2$, where $m$ is an integer number and $N_a$ the number of atoms per zigzag row. Only in the latter case AGNRs are metallic in the tight-binding model, the left two-thirds are semiconducting. \textit{Ab initio} calculations have shown that no truly metallic AGNR exist, but the gap remains very small.\cite{Cresti:2008} However, our system studied in this paper belongs to the ($3m+2$)-class and therefore shows metallicity in a simple nearest-neighbor tight-binding approximation, as can be nicely seen in Fig.\,\ref{fig:str}. We want to point out that our findings are qualitatively independent on the chosen class. The band gap oscillation discussed later in the text (see Fig.\,\ref{fig:gap}) is only shifted along the strain axis and is affected by defects in the same way.

By applying longitudinal uniaxial strain $\varepsilon_\parallel=\Delta L/L$ of up to 15\% one observes a transverse compression $\varepsilon_\perp=\Delta W/W$ of up to 1.7\%, leading to a positive Poisson's ratio $\nu=-\varepsilon_\perp/\varepsilon_\parallel$, which tends to decrease with increasing strain (see Fig.\,\ref{fig:stress-strain}). Its values are consistent with other ab-initio studies.\cite{Kudin:2001,Reddy:2006,Faccio:2009} For simplicity and due to the chosen modular model used here, we have neglected the edge effect,\cite{Sun:2008qy} i.e. stronger carbon bonds for edge atoms in AGNRs resulting in a slightly different geometry at the boundaries. 

To account for a more realistic contact model in an application point of view, the ribbon is side-contacted to metallic wide-band electrodes, in contrast to the widely used semi-infinite graphene leads. In the latter case the conductance turns out to be an integer multiple of the conductance quantum $G_0$, as the channels are transparent. This step function resembles the quantum upper-limit for the observed conductance and is therefore also shown in the conductance plots. For our model, the additional constraint in length leads to pronounced Fabry-P\'{e}rot-like oscillations of the transmission, see Fig.\,\ref{fig:str}. Those Fabry-P\'{e}rot oscillations can be useful in experiments as they allow to measure the length of the scattering region. On the other hand, weakly coupled contacts can lead to a different behaviour, the Coulomb blockade regime.\cite{Sols:2007,Stampfer:2008,Han:2007} We will not consider this in the present paper, as we assume a rather good contact.
As Palladium turns out to be one of the most favorable electrode material for carbon nanostructures,\cite{Nemec:2008,Nemec:2006} we have used the Palladium-specific coupling parameter $\gamma_{\mathrm{C-Pd}}=0.15$ eV for the wide-band coupling. This value for the Pd-C coupling has been proposed by Nemec \textit{et al.}\cite{Nemec:2008} by fitting the tight-binding Pd-C hopping integral to reproduce the electronic band structure of the hybridized metal-graphene system near the Fermi level.

\begin{figure}
\begin{center}
\epsfxsize=1\hsize
\epsfbox{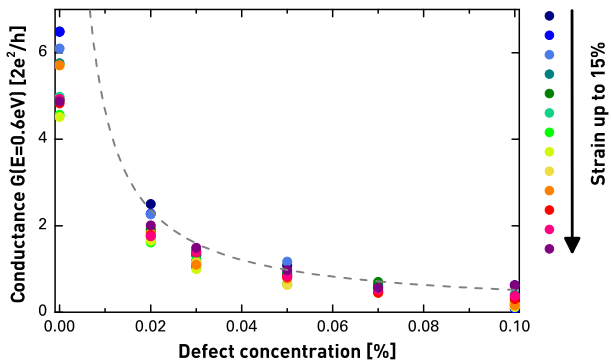}
\caption{(Color online) Conductance at fixed energy $E=0.6$ eV for various defect concentrations and strain values. An inverse scaling behaviour can be observed, i.e. $G(E=0.6\textrm{ eV}) \propto \rho_{\mathrm{Defects}}^{-1}$, due to scattering and localization.}
\label{fig:Tvsdef}
\end{center}
\end{figure}

When incorporating a finite defect concentration in the unstrained ribbon, the conductance properties are dramatically changed,\cite{Haskins:2011} see Fig.\,\ref{fig:str}a. Transmission is strongly suppressed even for very small concentrations of 0.02\% and the graphene-like features disappear, i.e. conductance plateaus vanish, electron-hole symmetry is broken and metallicity is destroyed. This effect gets enhanced by increasing the amount of defects, as can be seen in Fig.\,\ref{fig:Tvsdef}. One observes an inverse dependency of the conductance on defect concentration.

The picture is completely different if one considers the effect of tensile uniaxial strain in the ideal ribbon. The conductance is only slightly suppressed for high energies, but symmetry and plateaus remain to be visible, see Fig.\,\ref{fig:str}b. The largest effect is observed near the Fermi level, as strain can induce a band gap in metallic AGNRs, or change an existent band gap in semiconducting ribbons in a non-monotonous way. This peculiarity, exhibiting a zigzag feature has been studied extensively before and is now well understood.\cite{Lu:2010fj,Faccio:2009,Sun:2008qy,Li:2010,Pereira:2009,Rosenkranz:2011,Ni:2008rt,Lu:2012,Poetschke:2010,Erdogan:2011} 

The open question was how this observed behaviour is altered by vacancy defects, as they are present in experimental realizations. It is important to note, that the conductance gap near the Fermi level opened by defects is not a spectral band gap, as it is the case for mechanical deformation. This gap is induced by enhanced scattering and localization at the defected sites, as can be seen in the density of states (DOS). Thus we will use the word pseudo gap instead of band gap in the remaining part of the present work to account for this circumstance.

Figure \ref{fig:gap} shows, how strain and defects affect the pseudo gap. In the case for no disorder, the linear oscillatory dependency of the band gap can be nicely observed. Including a small defect concentration, this zigzag feature is still preserved, but due to localization of low-energy states the gap can not be re-closed completely by applying further strain. The same holds true \textit{vice versa}, an existing band gap generated by strain can not be closed by disorder. Hence, the inclusion of defects provides an additional offset of the energy gap but also leads to strong suppression of the conductance. The oscillation is expected to disappear for high concentrations of about $0.1\%$ or higher, as more and more states will be localized, but this regime is not interesting for applications because of its low conductivity. However, low disorder shows little effect on the qualitative behaviour of the energy gap, proving that mechanical strain is a real tool for effective band gap engineering.

\section{Conclusions}
\label{sec:conclusions}
In summary, we presented in this work a novel approach to set up geometries of large scale strained graphene nanoribbons with structural defects to account for more realistic system configurations and showed how one can effectively investigate its transport properties. The inclusion of vacancy defects shows a strong suppression of transmission due to scattering and localization, whereas mechanical strain affects largely only the conductance near the Fermi level by modifying the spectral band gap. The known oscillatory behaviour of the energy gap induced by uniaxial strain has been reproduced. Combining the effects of strain and defects, we revealed how a finite concentration of several vacancy defects affects this peculiarity. As it turns out, the zigzag feature is preserved for low disorder, which leaves mechanical strain as a valuable tool to tune the energy gap in non-ideal GNRs, applicable for future nanodevices.

\section*{Acknowledgments}
We acknowledge fruitful discussions with Arezoo Dianat and Daijiro Nozaki.

This work was supported by the Deutsche Forschungsgemeinschaft within the priority program “Nanostructured Thermoelectrics” (Grant No. SPP 1386) and from the German Excellence Initiative via the Cluster of Excellence 1056 “Center for Advancing Electronics Dresden” (cfAED). One of us (T. L.) thanks the national scholarship program Deutschlandstipendium. Computing time provided by the ZIH at the Dresden University of Technology is acknowledged.

\nocite{Cresti:2008,Erdogan:2011}
\bibliography{bibliography.bib}

\begin{thebibliography}{32}
\expandafter\ifx\csname natexlab\endcsname\relax\def\natexlab#1{#1}\fi
\expandafter\ifx\csname bibnamefont\endcsname\relax
  \def\bibnamefont#1{#1}\fi
\expandafter\ifx\csname bibfnamefont\endcsname\relax
  \def\bibfnamefont#1{#1}\fi
\expandafter\ifx\csname citenamefont\endcsname\relax
  \def\citenamefont#1{#1}\fi
\expandafter\ifx\csname url\endcsname\relax
  \def\url#1{\texttt{#1}}\fi
\expandafter\ifx\csname urlprefix\endcsname\relax\def\urlprefix{URL }\fi
\providecommand{\bibinfo}[2]{#2}
\providecommand{\eprint}[2][]{\url{#2}}

\bibitem[{\citenamefont{Novoselov et~al.}(2004)\citenamefont{Novoselov, Geim,
  Morozov, Jiang, Zhang, Dubonos, Grigorieva, and Firsov}}]{Novoselov:2004}
\bibinfo{author}{\bibfnamefont{K.~S.} \bibnamefont{Novoselov}},
  \bibinfo{author}{\bibfnamefont{A.~K.} \bibnamefont{Geim}},
  \bibinfo{author}{\bibfnamefont{S.~V.} \bibnamefont{Morozov}},
  \bibinfo{author}{\bibfnamefont{D.}~\bibnamefont{Jiang}},
  \bibinfo{author}{\bibfnamefont{Y.}~\bibnamefont{Zhang}},
  \bibinfo{author}{\bibfnamefont{S.~V.} \bibnamefont{Dubonos}},
  \bibinfo{author}{\bibfnamefont{I.~V.} \bibnamefont{Grigorieva}},
  \bibnamefont{and} \bibinfo{author}{\bibfnamefont{A.~A.}
  \bibnamefont{Firsov}}, \bibinfo{journal}{Science}
  \textbf{\bibinfo{volume}{306}}, \bibinfo{pages}{666} (\bibinfo{year}{2004}).

\bibitem[{\citenamefont{Lu and Guo}(2010)}]{Lu:2010fj}
\bibinfo{author}{\bibfnamefont{Y.}~\bibnamefont{Lu}} \bibnamefont{and}
  \bibinfo{author}{\bibfnamefont{J.}~\bibnamefont{Guo}}, \bibinfo{journal}{Nano
  Res.} \textbf{\bibinfo{volume}{3}}, \bibinfo{pages}{189}
  (\bibinfo{year}{2010}).

\bibitem[{\citenamefont{Faccio et~al.}(2009)\citenamefont{Faccio, Denis, Pardo,
  Goyenola, and Mombru}}]{Faccio:2009}
\bibinfo{author}{\bibfnamefont{R.}~\bibnamefont{Faccio}},
  \bibinfo{author}{\bibfnamefont{P.~A.} \bibnamefont{Denis}},
  \bibinfo{author}{\bibfnamefont{H.}~\bibnamefont{Pardo}},
  \bibinfo{author}{\bibfnamefont{C.}~\bibnamefont{Goyenola}}, \bibnamefont{and}
  \bibinfo{author}{\bibfnamefont{A.~W.} \bibnamefont{Mombru}},
  \bibinfo{journal}{J. Phys.: Condens. Matter} \textbf{\bibinfo{volume}{21}},
  \bibinfo{pages}{285304} (\bibinfo{year}{2009}).

\bibitem[{\citenamefont{Sun et~al.}(2008)\citenamefont{Sun, Li, Ren, Su, Shi,
  and Yang}}]{Sun:2008qy}
\bibinfo{author}{\bibfnamefont{L.}~\bibnamefont{Sun}},
  \bibinfo{author}{\bibfnamefont{Q.~X.} \bibnamefont{Li}},
  \bibinfo{author}{\bibfnamefont{H.}~\bibnamefont{Ren}},
  \bibinfo{author}{\bibfnamefont{H.~B.} \bibnamefont{Su}},
  \bibinfo{author}{\bibfnamefont{Q.~W.} \bibnamefont{Shi}}, \bibnamefont{and}
  \bibinfo{author}{\bibfnamefont{J.~L.} \bibnamefont{Yang}},
  \bibinfo{journal}{J. Chem. Phys} \textbf{\bibinfo{volume}{129}},
  \bibinfo{pages}{074704} (\bibinfo{year}{2008}).

\bibitem[{\citenamefont{Li et~al.}(2010)\citenamefont{Li, Jiang, Liu, and
  Liu}}]{Li:2010}
\bibinfo{author}{\bibfnamefont{Y.}~\bibnamefont{Li}},
  \bibinfo{author}{\bibfnamefont{X.~W.} \bibnamefont{Jiang}},
  \bibinfo{author}{\bibfnamefont{Z.~F.} \bibnamefont{Liu}}, \bibnamefont{and}
  \bibinfo{author}{\bibfnamefont{Z.~R.} \bibnamefont{Liu}},
  \bibinfo{journal}{Nano Res.} \textbf{\bibinfo{volume}{3}},
  \bibinfo{pages}{545} (\bibinfo{year}{2010}).

\bibitem[{\citenamefont{Pereira et~al.}(2009)\citenamefont{Pereira, Neto, and
  Peres}}]{Pereira:2009}
\bibinfo{author}{\bibfnamefont{V.~M.} \bibnamefont{Pereira}},
  \bibinfo{author}{\bibfnamefont{A.~H.~C.} \bibnamefont{Neto}},
  \bibnamefont{and} \bibinfo{author}{\bibfnamefont{N.~M.~R.}
  \bibnamefont{Peres}}, \bibinfo{journal}{Phys. Rev. B}
  \textbf{\bibinfo{volume}{80}}, \bibinfo{pages}{045401}
  (\bibinfo{year}{2009}).

\bibitem[{\citenamefont{Rosenkranz et~al.}(2011)\citenamefont{Rosenkranz, Mohr,
  and Thomsen}}]{Rosenkranz:2011}
\bibinfo{author}{\bibfnamefont{N.}~\bibnamefont{Rosenkranz}},
  \bibinfo{author}{\bibfnamefont{M.}~\bibnamefont{Mohr}}, \bibnamefont{and}
  \bibinfo{author}{\bibfnamefont{C.}~\bibnamefont{Thomsen}},
  \bibinfo{journal}{Ann. Phys. (Berlin)} \textbf{\bibinfo{volume}{523}},
  \bibinfo{pages}{137} (\bibinfo{year}{2011}).

\bibitem[{\citenamefont{Ni et~al.}(2008)\citenamefont{Ni, Yu, Lu, Wang, Feng,
  and Shen}}]{Ni:2008rt}
\bibinfo{author}{\bibfnamefont{Z.~H.} \bibnamefont{Ni}},
  \bibinfo{author}{\bibfnamefont{T.}~\bibnamefont{Yu}},
  \bibinfo{author}{\bibfnamefont{Y.~H.} \bibnamefont{Lu}},
  \bibinfo{author}{\bibfnamefont{Y.~Y.} \bibnamefont{Wang}},
  \bibinfo{author}{\bibfnamefont{Y.~P.} \bibnamefont{Feng}}, \bibnamefont{and}
  \bibinfo{author}{\bibfnamefont{Z.~X.} \bibnamefont{Shen}},
  \bibinfo{journal}{ACS Nano} \textbf{\bibinfo{volume}{2}},
  \bibinfo{pages}{2301} (\bibinfo{year}{2008}).

\bibitem[{\citenamefont{Lu et~al.}(2012)\citenamefont{Lu, Liao, Wang, and
  Zheng}}]{Lu:2012}
\bibinfo{author}{\bibfnamefont{T.~Y.} \bibnamefont{Lu}},
  \bibinfo{author}{\bibfnamefont{X.~X.} \bibnamefont{Liao}},
  \bibinfo{author}{\bibfnamefont{H.~Q.} \bibnamefont{Wang}}, \bibnamefont{and}
  \bibinfo{author}{\bibfnamefont{J.~C.} \bibnamefont{Zheng}},
  \bibinfo{journal}{J. Mater. Chem.} \textbf{\bibinfo{volume}{22}},
  \bibinfo{pages}{10062} (\bibinfo{year}{2012}).

\bibitem[{\citenamefont{Poetschke et~al.}(2010)\citenamefont{Poetschke, Rocha,
  Torres, Roche, and Cuniberti}}]{Poetschke:2010}
\bibinfo{author}{\bibfnamefont{M.}~\bibnamefont{Poetschke}},
  \bibinfo{author}{\bibfnamefont{C.~G.} \bibnamefont{Rocha}},
  \bibinfo{author}{\bibfnamefont{L.~E. F.~F.} \bibnamefont{Torres}},
  \bibinfo{author}{\bibfnamefont{S.}~\bibnamefont{Roche}}, \bibnamefont{and}
  \bibinfo{author}{\bibfnamefont{G.}~\bibnamefont{Cuniberti}},
  \bibinfo{journal}{Phys. Rev. B} \textbf{\bibinfo{volume}{81}},
  \bibinfo{pages}{193404} (\bibinfo{year}{2010}).

\bibitem[{\citenamefont{Erdogan et~al.}(2011)\citenamefont{Erdogan, Popov,
  Rocha, Cuniberti, Roche, and Seifert}}]{Erdogan:2011}
\bibinfo{author}{\bibfnamefont{E.}~\bibnamefont{Erdogan}},
  \bibinfo{author}{\bibfnamefont{I.}~\bibnamefont{Popov}},
  \bibinfo{author}{\bibfnamefont{C.~G.} \bibnamefont{Rocha}},
  \bibinfo{author}{\bibfnamefont{G.}~\bibnamefont{Cuniberti}},
  \bibinfo{author}{\bibfnamefont{S.}~\bibnamefont{Roche}}, \bibnamefont{and}
  \bibinfo{author}{\bibfnamefont{G.}~\bibnamefont{Seifert}},
  \bibinfo{journal}{Phys. Rev. B} \textbf{\bibinfo{volume}{83}},
  \bibinfo{pages}{041401} (\bibinfo{year}{2011}).

\bibitem[{\citenamefont{Yang and Han}(2000)}]{Yang:2000}
\bibinfo{author}{\bibfnamefont{L.}~\bibnamefont{Yang}} \bibnamefont{and}
  \bibinfo{author}{\bibfnamefont{J.}~\bibnamefont{Han}},
  \bibinfo{journal}{Phys. Rev. Lett.} \textbf{\bibinfo{volume}{85}},
  \bibinfo{pages}{154} (\bibinfo{year}{2000}).

\bibitem[{\citenamefont{Soler et~al.}(2002)\citenamefont{Soler, Artacho, Gale,
  Garcia, Junquera, Ordejon, and Sanchez-Portal}}]{Soler:2002}
\bibinfo{author}{\bibfnamefont{J.~M.} \bibnamefont{Soler}},
  \bibinfo{author}{\bibfnamefont{E.}~\bibnamefont{Artacho}},
  \bibinfo{author}{\bibfnamefont{J.~D.} \bibnamefont{Gale}},
  \bibinfo{author}{\bibfnamefont{A.}~\bibnamefont{Garcia}},
  \bibinfo{author}{\bibfnamefont{J.}~\bibnamefont{Junquera}},
  \bibinfo{author}{\bibfnamefont{P.}~\bibnamefont{Ordejon}}, \bibnamefont{and}
  \bibinfo{author}{\bibfnamefont{D.}~\bibnamefont{Sanchez-Portal}},
  \bibinfo{journal}{J. Phys.: Condens. Matter} \textbf{\bibinfo{volume}{14}},
  \bibinfo{pages}{2745} (\bibinfo{year}{2002}).

\bibitem[{\citenamefont{Meyer et~al.}(2008)\citenamefont{Meyer, Kisielowski,
  Erni, Rossell, Crommie, and Zettl}}]{Meyer:2008}
\bibinfo{author}{\bibfnamefont{J.~C.} \bibnamefont{Meyer}},
  \bibinfo{author}{\bibfnamefont{C.}~\bibnamefont{Kisielowski}},
  \bibinfo{author}{\bibfnamefont{R.}~\bibnamefont{Erni}},
  \bibinfo{author}{\bibfnamefont{M.~D.} \bibnamefont{Rossell}},
  \bibinfo{author}{\bibfnamefont{M.~F.} \bibnamefont{Crommie}},
  \bibnamefont{and} \bibinfo{author}{\bibfnamefont{A.}~\bibnamefont{Zettl}},
  \bibinfo{journal}{Nano Lett.} \textbf{\bibinfo{volume}{8}},
  \bibinfo{pages}{3582} (\bibinfo{year}{2008}).

\bibitem[{\citenamefont{Dietl}(University of Karlsruhe, 2009)}]{Dietl}
\bibinfo{author}{\bibfnamefont{P.}~\bibnamefont{Dietl}},
  \bibinfo{journal}{Diploma thesis}  (\bibinfo{year}{University of Karlsruhe,
  2009}).

\bibitem[{\citenamefont{Castro~Neto et~al.}(2009)\citenamefont{Castro~Neto,
  Guinea, Peres, Novoselov, and Geim}}]{Castro-Neto:2009}
\bibinfo{author}{\bibfnamefont{A.~H.} \bibnamefont{Castro~Neto}},
  \bibinfo{author}{\bibfnamefont{F.}~\bibnamefont{Guinea}},
  \bibinfo{author}{\bibfnamefont{N.~M.~R.} \bibnamefont{Peres}},
  \bibinfo{author}{\bibfnamefont{K.~S.} \bibnamefont{Novoselov}},
  \bibnamefont{and} \bibinfo{author}{\bibfnamefont{A.~K.} \bibnamefont{Geim}},
  \bibinfo{journal}{Rev. Mod. Phys.} \textbf{\bibinfo{volume}{81}},
  \bibinfo{pages}{109} (\bibinfo{year}{2009}).

\bibitem[{\citenamefont{Datta}(1995)}]{Datta95book}
\bibinfo{author}{\bibfnamefont{S.}~\bibnamefont{Datta}},
  \emph{\bibinfo{title}{Electronic Transport in Mesoscopic Systems}}
  (\bibinfo{publisher}{Cambridge University Press, Cambridge},
  \bibinfo{year}{1995}).

\bibitem[{\citenamefont{Ferry et~al.}(2009)\citenamefont{Ferry, Goodnick, and
  Bird}}]{Ferry:2009}
\bibinfo{author}{\bibfnamefont{D.~K.} \bibnamefont{Ferry}},
  \bibinfo{author}{\bibfnamefont{S.~M.} \bibnamefont{Goodnick}},
  \bibnamefont{and} \bibinfo{author}{\bibfnamefont{J.}~\bibnamefont{Bird}},
  \emph{\bibinfo{title}{Transport in nanostructures}}
  (\bibinfo{publisher}{Cambridge University Press},
  \bibinfo{address}{Cambridge}, \bibinfo{year}{2009}), \bibinfo{edition}{2nd}
  ed.

\bibitem[{\citenamefont{Cuevas and Scheer}(2010)}]{Cuevas10book}
\bibinfo{author}{\bibfnamefont{J.~C.} \bibnamefont{Cuevas}} \bibnamefont{and}
  \bibinfo{author}{\bibfnamefont{E.}~\bibnamefont{Scheer}},
  \emph{\bibinfo{title}{Molecular electronics: An Introduction to Theory and
  Experiment}} (\bibinfo{publisher}{World Scientific, Singapore},
  \bibinfo{year}{2010}).

\bibitem[{\citenamefont{Ryndyk et~al.}(2009)\citenamefont{Ryndyk,
  Guti\'{e}rrez, Song, and Cuniberti}}]{Ryndyk09inbook}
\bibinfo{author}{\bibfnamefont{D.~A.} \bibnamefont{Ryndyk}},
  \bibinfo{author}{\bibfnamefont{R.}~\bibnamefont{Guti\'{e}rrez}},
  \bibinfo{author}{\bibfnamefont{B.}~\bibnamefont{Song}}, \bibnamefont{and}
  \bibinfo{author}{\bibfnamefont{G.}~\bibnamefont{Cuniberti}},
  \emph{\bibinfo{title}{Energy Flow Dynamics in Biomaterial Systems}}
  (\bibinfo{publisher}{Springer, Berlin (arXiv:0805.0628)},
  \bibinfo{year}{2009}), chap. \bibinfo{chapter}{Green function techniques in
  the treatment of quantum transport at the molecular scale}, p.
  \bibinfo{pages}{213}.

\bibitem[{\citenamefont{MacKinnon}(1985)}]{MacKinnon:1985}
\bibinfo{author}{\bibfnamefont{A.}~\bibnamefont{MacKinnon}},
  \bibinfo{journal}{EPJ B} \textbf{\bibinfo{volume}{59}}, \bibinfo{pages}{385}
  (\bibinfo{year}{1985}).

\bibitem[{\citenamefont{Thouless and Kirkpatrick}(1981)}]{Thouless:1981a}
\bibinfo{author}{\bibfnamefont{D.~J.} \bibnamefont{Thouless}} \bibnamefont{and}
  \bibinfo{author}{\bibfnamefont{S.}~\bibnamefont{Kirkpatrick}},
  \bibinfo{journal}{Journal of Physics C: Solid State Physics}
  \textbf{\bibinfo{volume}{14}}, \bibinfo{pages}{235} (\bibinfo{year}{1981}).

\bibitem[{\citenamefont{Lee and Fisher}(1981)}]{LEE:1981}
\bibinfo{author}{\bibfnamefont{P.~A.} \bibnamefont{Lee}} \bibnamefont{and}
  \bibinfo{author}{\bibfnamefont{D.~S.} \bibnamefont{Fisher}},
  \bibinfo{journal}{Phys. Rev. Lett.} \textbf{\bibinfo{volume}{47}},
  \bibinfo{pages}{882} (\bibinfo{year}{1981}).

\bibitem[{\citenamefont{Kudin et~al.}(2001)\citenamefont{Kudin, Scuseria, and
  Yakobson}}]{Kudin:2001}
\bibinfo{author}{\bibfnamefont{K.~N.} \bibnamefont{Kudin}},
  \bibinfo{author}{\bibfnamefont{G.~E.} \bibnamefont{Scuseria}},
  \bibnamefont{and} \bibinfo{author}{\bibfnamefont{B.~I.}
  \bibnamefont{Yakobson}}, \bibinfo{journal}{Phys. Rev. B}
  \textbf{\bibinfo{volume}{64}}, \bibinfo{pages}{235406}
  (\bibinfo{year}{2001}).

\bibitem[{\citenamefont{Reddy et~al.}(2006)\citenamefont{Reddy, Rajendran, and
  Liew}}]{Reddy:2006}
\bibinfo{author}{\bibfnamefont{C.~D.} \bibnamefont{Reddy}},
  \bibinfo{author}{\bibfnamefont{S.}~\bibnamefont{Rajendran}},
  \bibnamefont{and} \bibinfo{author}{\bibfnamefont{K.~M.} \bibnamefont{Liew}},
  \bibinfo{journal}{Nanotechnology} \textbf{\bibinfo{volume}{17}},
  \bibinfo{pages}{864} (\bibinfo{year}{2006}).

\bibitem[{\citenamefont{Cresti et~al.}(2008)\citenamefont{Cresti, Nemec, Biel,
  Niebler, Triozon, Cuniberti, and Roche}}]{Cresti:2008}
\bibinfo{author}{\bibfnamefont{A.}~\bibnamefont{Cresti}},
  \bibinfo{author}{\bibfnamefont{N.}~\bibnamefont{Nemec}},
  \bibinfo{author}{\bibfnamefont{B.}~\bibnamefont{Biel}},
  \bibinfo{author}{\bibfnamefont{G.}~\bibnamefont{Niebler}},
  \bibinfo{author}{\bibfnamefont{F.}~\bibnamefont{Triozon}},
  \bibinfo{author}{\bibfnamefont{G.}~\bibnamefont{Cuniberti}},
  \bibnamefont{and} \bibinfo{author}{\bibfnamefont{S.}~\bibnamefont{Roche}},
  \bibinfo{journal}{Nano Res.} \textbf{\bibinfo{volume}{1}},
  \bibinfo{pages}{361} (\bibinfo{year}{2008}).

\bibitem[{\citenamefont{Sols et~al.}(2007)\citenamefont{Sols, Guinea, and
  Neto}}]{Sols:2007}
\bibinfo{author}{\bibfnamefont{F.}~\bibnamefont{Sols}},
  \bibinfo{author}{\bibfnamefont{F.}~\bibnamefont{Guinea}}, \bibnamefont{and}
  \bibinfo{author}{\bibfnamefont{A.~H.~C.} \bibnamefont{Neto}},
  \bibinfo{journal}{Phys. Rev. Lett.} \textbf{\bibinfo{volume}{99}},
  \bibinfo{pages}{166803} (\bibinfo{year}{2007}).

\bibitem[{\citenamefont{Stampfer et~al.}(2008)\citenamefont{Stampfer,
  Guttinger, Molitor, Graf, Ihn, and Ensslin}}]{Stampfer:2008}
\bibinfo{author}{\bibfnamefont{C.}~\bibnamefont{Stampfer}},
  \bibinfo{author}{\bibfnamefont{J.}~\bibnamefont{Guttinger}},
  \bibinfo{author}{\bibfnamefont{F.}~\bibnamefont{Molitor}},
  \bibinfo{author}{\bibfnamefont{D.}~\bibnamefont{Graf}},
  \bibinfo{author}{\bibfnamefont{T.}~\bibnamefont{Ihn}}, \bibnamefont{and}
  \bibinfo{author}{\bibfnamefont{K.}~\bibnamefont{Ensslin}},
  \bibinfo{journal}{Appl. Phys. Lett.} \textbf{\bibinfo{volume}{92}},
  \bibinfo{pages}{012102} (\bibinfo{year}{2008}).

\bibitem[{\citenamefont{Han et~al.}(2007)\citenamefont{Han, Ozyilmaz, Zhang,
  and Kim}}]{Han:2007}
\bibinfo{author}{\bibfnamefont{M.~Y.} \bibnamefont{Han}},
  \bibinfo{author}{\bibfnamefont{B.}~\bibnamefont{Ozyilmaz}},
  \bibinfo{author}{\bibfnamefont{Y.~B.} \bibnamefont{Zhang}}, \bibnamefont{and}
  \bibinfo{author}{\bibfnamefont{P.}~\bibnamefont{Kim}},
  \bibinfo{journal}{Phys. Rev. Lett.} \textbf{\bibinfo{volume}{98}},
  \bibinfo{pages}{206805} (\bibinfo{year}{2007}).

\bibitem[{\citenamefont{Nemec et~al.}(2008)\citenamefont{Nemec, Tomanek, and
  Cuniberti}}]{Nemec:2008}
\bibinfo{author}{\bibfnamefont{N.}~\bibnamefont{Nemec}},
  \bibinfo{author}{\bibfnamefont{D.}~\bibnamefont{Tomanek}}, \bibnamefont{and}
  \bibinfo{author}{\bibfnamefont{G.}~\bibnamefont{Cuniberti}},
  \bibinfo{journal}{Phys. Rev. B} \textbf{\bibinfo{volume}{77}},
  \bibinfo{pages}{125420} (\bibinfo{year}{2008}).

\bibitem[{\citenamefont{Nemec et~al.}(2006)\citenamefont{Nemec, Tomanek, and
  Cuniberti}}]{Nemec:2006}
\bibinfo{author}{\bibfnamefont{N.}~\bibnamefont{Nemec}},
  \bibinfo{author}{\bibfnamefont{D.}~\bibnamefont{Tomanek}}, \bibnamefont{and}
  \bibinfo{author}{\bibfnamefont{G.}~\bibnamefont{Cuniberti}},
  \bibinfo{journal}{Phys. Rev. Lett.} \textbf{\bibinfo{volume}{96}},
  \bibinfo{pages}{076802} (\bibinfo{year}{2006}).

\bibitem[{\citenamefont{Haskins et~al.}(2011)\citenamefont{Haskins, Kinaci,
  Sevik, Sevincli, Cuniberti, and Cagin}}]{Haskins:2011}
\bibinfo{author}{\bibfnamefont{J.}~\bibnamefont{Haskins}},
  \bibinfo{author}{\bibfnamefont{A.}~\bibnamefont{Kinaci}},
  \bibinfo{author}{\bibfnamefont{C.}~\bibnamefont{Sevik}},
  \bibinfo{author}{\bibfnamefont{H.}~\bibnamefont{Sevincli}},
  \bibinfo{author}{\bibfnamefont{G.}~\bibnamefont{Cuniberti}},
  \bibnamefont{and} \bibinfo{author}{\bibfnamefont{T.}~\bibnamefont{Cagin}},
  \bibinfo{journal}{ACS Nano} \textbf{\bibinfo{volume}{5}},
  \bibinfo{pages}{3779} (\bibinfo{year}{2011}).

\end{thebibliography}

\end{document}